\documentclass[aps,prb,superscriptaddress,twocolumn,floatfix]{revtex4-2}
\usepackage{graphicx,color}
\usepackage{dcolumn}
\usepackage{bm}
\usepackage{stmaryrd}
\usepackage{latexsym}
\usepackage{amssymb}
\usepackage{amsfonts}
\usepackage{amsmath}
\usepackage{subfigure}
\usepackage{hyperref}
\usepackage{verbatim}
\usepackage{multirow}
\begin{document}
\preprint{APS/123-QED}

\title{Understanding the flat band in 1T-TaS$_2$ using a rotated basis}
\author{Li Cheng}
\affiliation{Shenzhen Geim Graphene Center (SGC), Tsinghua-Berkeley Shenzhen Institute (TBSI) and Tsinghua Shenzhen International Graduate School, Tsinghua University, Shenzhen 518055, China}
\affiliation{Institute for Advanced Study, Tsinghua University, Beijing 100084, China}
\author{Xuanyu Long}
\affiliation{Institute for Advanced Study, Tsinghua University, Beijing 100084, China}
\author{Xiaobin Chen}
\affiliation{School of Science, State Key Laboratory on Tunable laser Technology and Ministry of Industry and Information Technology Key Lab of
Micro-Nano Optoelectronic Information System, Harbin Institute of Technology, Shenzhen, Shenzhen 518055, China}
\author{Xiaolong Zou}
\email{xlzou@sz.tsinghua.edu.cn}
\affiliation{Shenzhen Geim Graphene Center (SGC), Tsinghua-Berkeley Shenzhen Institute (TBSI) and Tsinghua Shenzhen International Graduate School, Tsinghua University, Shenzhen 518055, China}
\author{Zheng Liu}
\email{zheng-liu@tsinghua.edu.cn}
\affiliation{Institute for Advanced Study, Tsinghua University, Beijing 100084, China}
\date{\today}

\begin{abstract}
Electronic flat bands serve as a unique platform to achieve strongly-correlated phases. The emergence of a flat band around the Fermi level in 1T-TaS$_2$ in accompany with the development of a $\sqrt{13}\times\sqrt{13}$ charge density wave (CDW) superlattice has long been noticed experimentally, but a transparent theoretical understanding remains elusive. We show that without CDW, the primary feature of the $1\times1$ bands can be fitted by a simple trigonometric function, and physically understood by choosing a rotated $\tilde{t}_{2g}$ basis with the principle axes aligning to the tilted TaS$_6$ octahedron. Using this basis, we trace the band evolution in the $\sqrt{13}\times\sqrt{13}$ superlattice by progressively including different CDW effects. We point out that CDW strongly rehybridizes the three $\tilde{t}_{2g}$ orbitals, which leads to the formation of a well-localized molecular orbital and spawns the flat band. 
\end{abstract}

\maketitle

The layered transition-metal dichalcogenide (TMD) 1T-TaS$_2$ attracts revived interest recently due to its intriguing electronic properties intertwined with the charge density wave (CDW) order\cite{PRB_2015_cho,lee2017,2017high,Natphys_Geck_2015,npj2020,2018stacking,prl2019lee,NC2020Butler}. Despite debates on the nature of the low-temperature insulating and paramagnetic state, as well as the role of interlayer coupling, it is becoming more and more clear experimentally~\cite{NC2020Zhang,PRL2021} that a nearly flat band forms around the Fermi level in accompany with the development of CDW, which represents one of the most prominent features of this system.

Intuitively, it is not surprising that the long CDW period ($\sqrt{13}\times\sqrt{13}$ under around 200 K~\cite{APS_1975,Scruby1975,PMB_1979,Rossnagel_2011}) tends to reduce the overall band width by band folding. However, similar to the situation of twisted bilayer graphene at the magic angle~\cite{PNAS12233}, the real puzzle is that an isolated band around the Fermi level has a particularly  small band width in comparison with the other bands of the same system. An earlier tight-binding (TB)  study~\cite{smith1985band} carefully considered the standard power-law modulation of the Slater-Koster (SK) parameters~\cite{SK} associated with the CDW-induced bond length change, but only found a ``spaghetti'' of intertwisted CDW subbands around the Fermi level. It was later shown that an isolated flat band could be split out by adding spin-orbit coupling (SOC)~\cite{RossnagelPRB2006}. However, more recent first-principles calculations 
indicate that SOC is not a necessary ingredient - the flat band in 1T-TaS$_2$ can be reproduced within the standard non-relativistic framework of density functional theory (DFT)\cite{bulk2017,TMDCs2020}, and the SOC effect on the $\sqrt{13}\times\sqrt{13}$ bands is negligible~\cite{PRB2014Darancet,PRB2018SOC}. To date, answers to several fundamental questions remain elusive, such as ``what is the key mechanism to create such a singular band structure?'' and ``why does it happen to locate at the Fermi level?''

In this Letter, we aim to provide a transparent theoretical understanding by combining TB modeling and first-principles Wannier function (WF) analysis. We find that a rotated $\tilde{t}_{2g}$ basis with the principle axes aligning to the local Ta-S bonds dramatically simplifies the analysis of band formation, providing a heuristic framework to understand the origin of the flat band.

\begin{figure*}
\centering
\includegraphics[width=18cm]{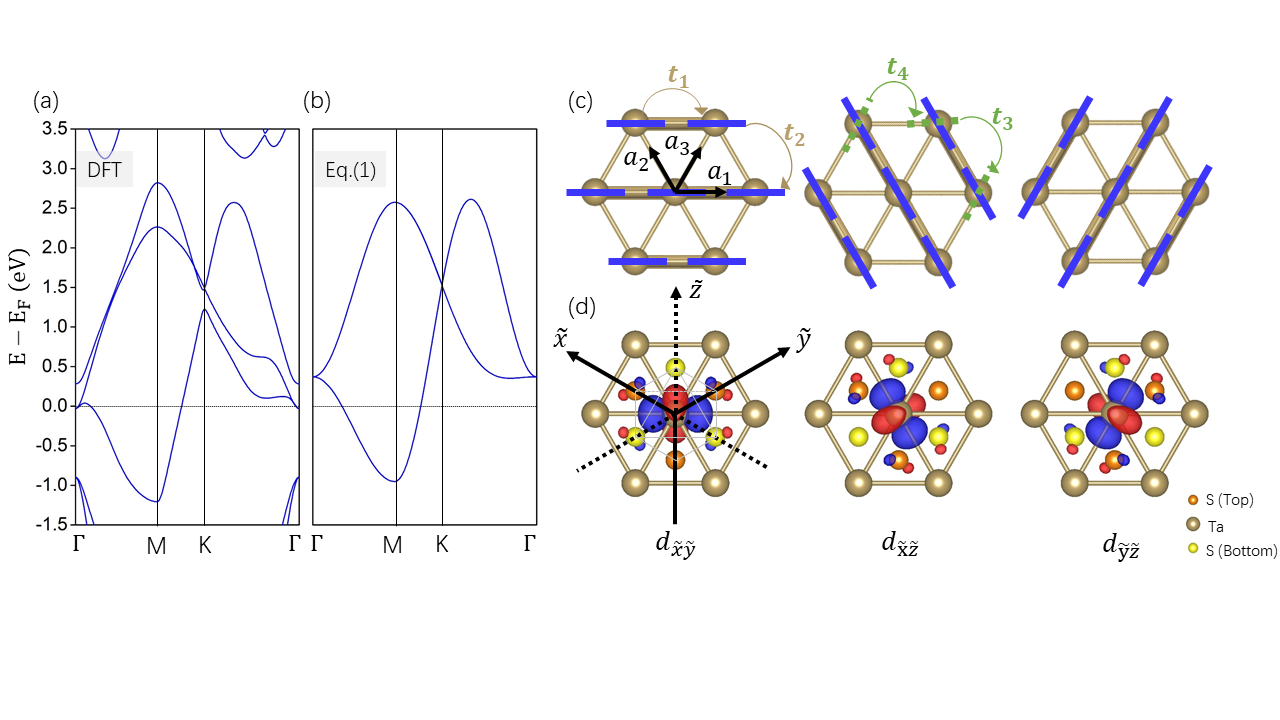}
\caption{(a) $1\times1$ band structure from DFT, and (b) a simple fitting using Eq. \ref{eq:1}; (c) The three orbital orientations and directional hoppings implied by Eq. (\ref{eq:1}). In addition to the major intra-orbital hoppings ($t_1$ and $t_2$), we also mark the minor $t_3$ and $t_4$ between two different types of orbitals as depicted by the green dashed lines; (d) The first-principles WFs associated with the three bands around the Fermi level constructed under the rotated axes $\tilde{x}\tilde{y}\tilde{z}$.}
\label{fig1}
\end{figure*}

\begin{figure*}
\centering
\includegraphics[width=18cm]{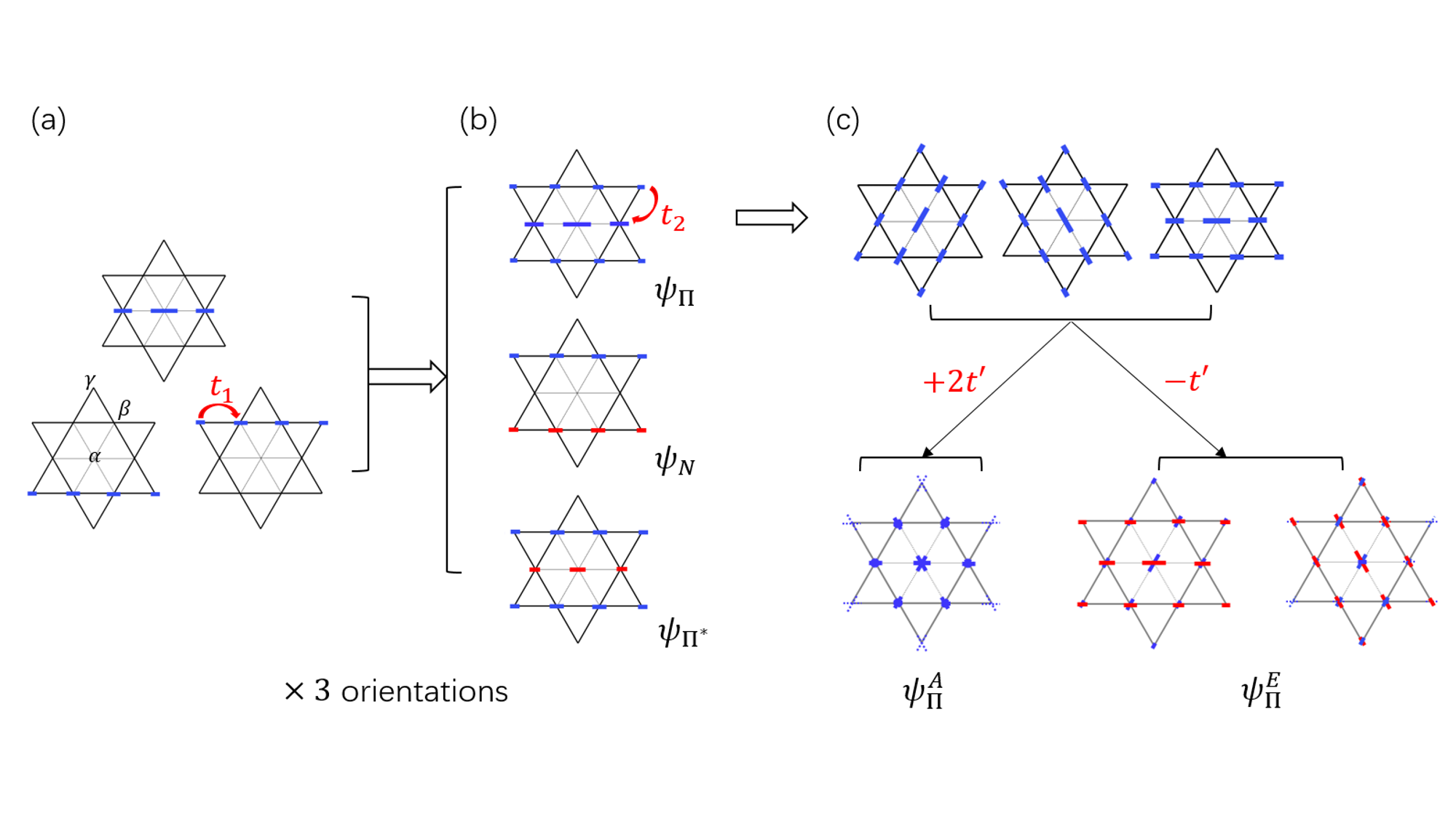}
\caption{Schematics of the molecular orbitals formed within an SD. The blue (red) short line at a Ta site represents the $\tilde{t}_{2g}$ component with a positive (negative) sign. The length of the line roughly reflects the weight.}
\label{fig2}
\end{figure*}

We first discuss the 1$\times$1 band structure. Figure \ref{fig1}(a) shows the first-principles result of a single layer. We note that the flat band is a intra-layer feature. The problem of inter-layer stacking and the resulted out-of-plane reconstruction is beyond the scope of the current work. Our first-principles calculation is performed using the Vienna \textit{ab initio} Simulation Package (VASP) \cite{kresse1996,kresse1996efficient}. The numerical setup follows our previous work on the same system~\cite{PRR2020}. 

To fit the first-principles bands, the previous TB analysis~\cite{smith1985band, RossnagelPRB2006} invoked six parameters - the nearest-neighbor (NN)  hopping integrals $dd\sigma$, $dd\pi$ and $dd\delta$ of the Ta 5d-orbitals and the onsite energies of $d_{z^2}$, $d_{xy/x^2-y^2}$ and $d_{yz/xz}$. Within this framework, the three bands shown in Fig. \ref{fig1}(a) correspond to the lower three eigen-levels of a 5$\times$5 hopping matrix. 

Interestingly, if neglecting band splittings, we noticed that the general features of the three bands could be fitted by a simple trigonometric function [Fig. \ref{fig1}(b)] with two parameters:
\begin{eqnarray}\label{eq:1}
\epsilon_{k} = 2t_1 \cos k_{1} + 2t_{2}(\cos k_2 + \cos k_3),
\end{eqnarray}
and by permutating $k_{1,2,3}$ to expand this single dispersion into a three-band multiplet, where $k_{i}$ = $\vec{k}\cdot\vec{a}_{i}$ and $\vec{a}_{i}$'s are the three NN vectors of a triangular lattice [see labels in Fig. \ref{fig1}(c)]. Physically, Eq. (\ref{eq:1}) corresponds to an orbital with strong bonding $t_1$ along  $\vec a_1$ and weak bonding $t_2$ along $\vec a_{2,3}$. Due to the three-fold rotational symmetry of the triangular lattice, there are two other equivalent orbital orientations. Figure \ref{fig1}(c) schematically draws the three orbital orientations. It suggests that by properly defining the principle axes of the 5$d$ orbitals, in a rotated new basis, the primary physics can be captured by three nearly independent orbitals with intra-orbital hoppings only, which will drastically simplify our analysis of the CDW effect. The inter-orbital hoppings do exist, as reflected by the extra band splittings in Fig. \ref{fig1}(a) in comparison to Fig. \ref{fig1}(b), which will be included later when the intra-orbital physics is fully understood.

To determine the rotated axes, we note that each Ta atom bonds with the six NN S atoms forming an octahedron, and the local octahedral crystal field naturally splits out a lower-energy $\tilde{t}_{2g}$ triplet consisting of $d_{\tilde{x}\tilde{y}}$, $d_{\tilde{x}\tilde{z}}$ and $d_{\tilde{y}\tilde{z}}$ orbitals. Here, to differentiate from the $x$, $y$, $z$-axes commonly defined for a two-dimensional triangular lattice, i.e. $x$-axis parallel to $\vec{a}_1$ and $z$-axis perpendicular to the plane, we use $\tilde{\square} $ to denote the rotated axes referring to the Ta-S bonds [c.f. Fig. 1(d)]. The three orbital orientations shown in Fig. 1(c) coincide with lobes of the $d_{\tilde{x}\tilde{y}}$, $d_{\tilde{x}\tilde{z}}$ and $d_{\tilde{y}\tilde{z}}$ orbitals. As a first-principles justification, we construct the maximally localized WFs~\cite{W90} within the three-band subspace. The minimization procedure automatically produces three well-localized WFs with the desired orientations [Fig. 1(d)]. Despite hybridizations with the NN S $p$-orbitals, we see that the $d_{\tilde{x}\tilde{y}}$, $d_{\tilde{x}\tilde{z}}$ and $d_{\tilde{y}\tilde{z}}$ orbitals represent good approximations to these WFs. 

It is not apparent that the inter-orbital hopping can be largely eliminated in the rotated basis. Besides the intra-orbital hoppings $t_1$ and $t_2$, there are two types of inter-orbital hoppings $t_3$ and $t_4$, as schematically shown in Fig. \ref{fig1}(c). The hopping parameters between the WFs can be readily obtained by Fourier transforming the first-principles band structure~\cite{W90}. The results (second row of Tab. I) show that $t_3$ and $t_4$ are indeed small. Clear physical insights can be obtained by switching back to the SK parameterization, where the $t$'s are linear combinations of $dd$  $\sigma$, $\pi$ and $\delta$ bonds. In the rotated basis, the relations are: $t_1=\frac{3}{4}dd\sigma+\frac{1}{4}dd\delta$, $t_2=\frac{1}{2}dd\pi+\frac{1}{2}dd\delta$, $t_3=\frac{1}{2}dd\pi-\frac{1}{2}dd\delta$, $t_4=0$, according to the standard SK table~\cite{SK}. Plugging in the $dd\sigma$, $dd\pi$ and $dd\delta$ values fitted previously~\cite{smith1985band}, the SK estimations are summarized in the first row of Tab. I, which are comparable to the WF results. We now see that the small magnitude of $t_3$ arises from a partial cancellation between $dd\pi$ and $dd\delta$, whereas $t_4$ vanishes for a symmetry reason. Specifically, for the $t_4$ hopping marked in the middle panel of Fig. \ref{fig1}(c) between a $d_{\tilde{y}\tilde{z}}$ orbital and a $d_{\tilde{x}\tilde{y}}$ orbital, the hopping path lies in the $\tilde{x}\tilde{y}$ plane, while the two orbitals have opposite parities with respect to this plane. Therefore, within the SK two-centre approximation, such a hopping process is strictly forbidden~\cite{SK}. 

We now proceed to consider the CDW effects. At low temperature, the CDW is known to freeze the structure into the so-called ``Star-of-David'' (SD) motifs with $\sqrt{13}\times\sqrt{13}$ periodicity~\cite{APS_1975,Scruby1975,PMB_1979,Rossnagel_2011}, which differentiates three inequivalent Ta sites, labeled as $\alpha$, $\beta$ and $\gamma$ in Fig. \ref{fig2}(a),  following the notation previously used in experiment~\cite{181Ta,2017high}. Naturally, such a CDW order strengthens the intra-SD hoppings, and weakens the inter-SD hoppings. Combining the condition $t_1> t_2 > t_{3,4}$, it is heuristic to first study the intra-orbital hybridization within one SD, then construct the molecular orbitals including the inter-orbital mixing,  and finally view the $\sqrt{13}\times\sqrt{13}$ bands as overlap and broadening of these molecular orbitals. A step-by-step analysis is as follows:

\textit{Step I}: Consider intra-SD $t_1$ only. The SD decouples into independent chains. The three lowest-energy chain states associated with the $d_{\tilde{x}\tilde{y}}$ orbital are depicted in Fig. \ref{fig2}(a). The negative sign of $t_1$ leads to energy gain by forming bonds.  The central $\beta$-$\alpha$-$\beta$ chain [top panel of Fig. \ref{fig2}(a)] has a higher energy than the two surrounding $\gamma$-$\beta$-$\beta$-$\gamma$ chains [bottom panels of Fig. \ref{fig2}(a)] due to a fewer number of $t_1$ bonds. 

\textit{Step II}: Add intra-SD $t_2$. The three lowest-energy chain states formed in \textit{Step I} are further mixed into a bonding state ($\psi_{\Pi}$), a non-bonding state ($\psi_{N}$) and an anti-bonding state ($\psi_{\Pi^*}$) [Fig. \ref{fig2}(b)]. The positive sign of $t_2$ raises (reduces) the energy of the bonding (anti-bonding) state.
$\psi_{\Pi}$ with the highest energy inherits the largest central chain component.

Putting the three orbital orientations together, we now have nine molecular orbitals in total. With spin degeneracy, $\psi_{\Pi^*}^{1,2,3}$ and $\psi_{N}^{1,2,3}$ accommodate 12 electrons. The last unpaired 5d Ta electron has to go to one of the $\psi_{\Pi}$ states. For this highest level, the splitting induced by the smaller $t_{3,4}$ becomes relevant.

\textit{Step III}: Add intra-SD $t_{3,4}$. The mixing of $\psi_{\Pi}^{1,2,3}$ is governed by the Hamiltonian $t'\sum_{n\neq m}|\psi_{\Pi}^n\rangle\langle\psi_{\Pi}^m|$, in which $t'$ is a linear combination of $t_{3,4}$ depending on the forms of $\psi_{\Pi}^{1,2,3}$. The eigenstates consist of a singlet $\psi_{\Pi}^A=\frac{1}{\sqrt{3}}(1,1,1)^T$, and a doublet $\psi_{\Pi}^{E,1}=\frac{1}{\sqrt{2}}(-1,0,1)^T$, $\psi_{\Pi}^{E,2}=\frac{1}{\sqrt{6}}(1,-2,1)^T$ [Fig. \ref{fig2}(c)].

Importantly, $\psi_{\Pi}^A$ represents a unique molecular orbital strongly localized at the SD center, which minimizes the inter-SD overlap making an exceptionally narrow band possible. We note that the orbital texture associated with the flat band as observed in experiment~\cite{qiao2017} indeed resembles $\psi_{\Pi}^A$. 

However, to ensure that an isolated $\psi_{\Pi}^A$ band appears at the Fermi level, the minimal requirements are: (i) the sign of $t'$ is negative, so the $\psi_{\Pi}^A$ level is lower in energy than the $\psi_{\Pi}^E$ level, and (2) the magnitude $t'$ is at least comparable to the inter-SD $t_1$, so after the molecular levels are broaden into energy bands, the $\psi_{\Pi}^A$ band will stay separated from the $\psi_{\Pi}^E$ bands as well as the other high-energy bands.

\begin{table}
\centering
\setlength\extrarowheight{8pt}
    \begin{tabular}{ccccc}
    \hline
         {(eV)}& t$_1$& t$_2$& t$_3$& t$_4$  \\
    \hline
        {SK}& -0.5& 0.2& 0.1& 0.0\\
        {WF}& -0.7& 0.2& 0.1& 0.0 \\
    \hline
    \end{tabular}
    \caption{Hopping parameters under the rotated basis for the $1\times1$ structure from SK parameterization and WF analysis. For the SK parameters, we adopt the same $dd\sigma$, $dd\pi$ and $dd\delta$ values fitted previously~\cite{smith1985band,RossnagelPRB2006}}
    \label{tab:noCDW}
\end{table}

\begin{table}
\centering
\setlength\extrarowheight{8pt}
    \begin{tabular}{ccccc}
    \hline
         {(eV)}& t$_1$& t$_2$& t$_3$& t$_4$  \\
    \hline
        {SK}& $[-0.7,-0.3]$& $[0.1,0.3]$& $[0.0, 0.1]$ & 0.0\\
        {WF}& $[-1.1,-0.2]$& $[0.1, 0.2]$& $[-0.2, 0.1]$ & $[-0.2, 0.0]$ \\
    \hline
    \end{tabular}
    \caption{Hopping parameters under the rotated basis for the CDW-distorted $\sqrt{13}\times\sqrt{13}$ structure. For the SK parameters, the variation range is estimated by scaling the $1\times1$ parameters with a $d^{-5}$ law as used previously~\cite{smith1985band,RossnagelPRB2006}, where $d$ is the Ta-Ta distance. The Ta-Ta distance is extracted from the fully-relaxed first-principles structure}
    \label{tab:CDW}
\end{table}

\begin{figure*}
\centering
\includegraphics[width=18cm]{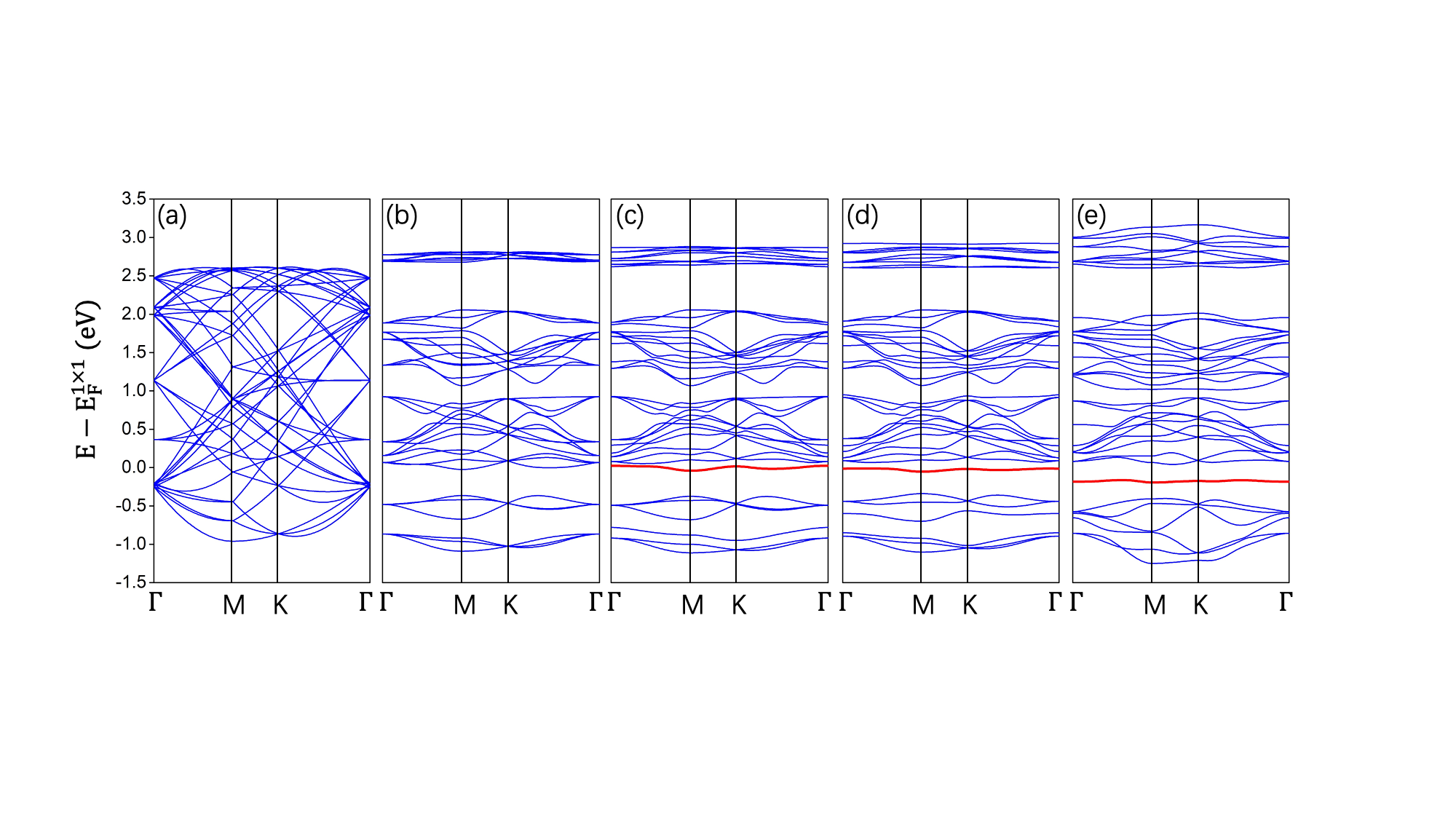}
\caption{Evolution of the $\sqrt{13}\times\sqrt{13}$ band structure calculation from a $39\times39$ hopping matrix. (a) $t_1=-0.7$ eV, $t_2=0.2$ eV, $t_{3,4}=0$; (b) $t_{1,2}$  according to the WF analysis for the CDW phase, $t_{3,4}=0$; (c) $t_{1,2}$ same as (b), plus the largest $|t_4|$; (d) $t_{1,2}$ same as (b), plus the largest $|t_4|$ and $|t_3|$; (d) full $t_{1,2,3,4}$ according to the WF analysis for the CDW phase. The flat band appearing in (c-e) is highlighted in red. }
\label{fig3}
\end{figure*}

We confirm that these requirements are met according to the analysis of first-principles WF hopping parameters. In the CDW structure, the hopping parameters vary from bond to bond. To obtain the hopping parameters in the $\sqrt{13}\times\sqrt{13}$ superlattice, 39 maximally-localized WFs are constructed. We summarize the variation range of $t_{1,2,3,4}$ in Tab. \ref{tab:CDW}. Note that large negative $t_3$ and $t_4$ emerge. The form of $\psi_{\Pi}^A$ can be formally written as $\sum_{I\in SD,n} w_{In}|d_{I,n}\rangle $, in which $I$ and $n$ denote the Ta site and the $\tilde{t}_{2g}$ orbital, respectively. Then, $t'=t_3\sum_{(I\alpha,J\beta)\in t_3}w_{I\alpha}w_{J\beta}+t_4\sum_{(I\alpha,J\beta)\in t_4}w_{I\alpha}w_{J\beta}$. As a bonding orbital, all the weights $w_{I,\alpha}$ have the same sign. Therefore, the large negative $t_3$ and $t_4$ cooperatively lead to a large negative $t'$.

To substantiate that such $t_3$ and $t_4$ are the key players to isolate a flat band, we plot the band evolution in Fig.  \ref{fig3} by selectively feeding different WF hopping parameters into a $39\times39$ matrix. As a benchmark, Fig. \ref{fig3}(a) uses only two parameters $t_1=-0.7$ eV and $t_2=0.2$ eV, as from the second row of Tab. \ref{tab:noCDW}, and the results are the simple folding of the $1\times1$ band structure.  Figure \ref{fig3}(b) includes the WF analysis results for $t_1$ and $t_2$ in the $\sqrt{13}\times\sqrt{13}$ structure, but keeps $t_3$ and $t_4$ zero.  The simply folded bands are split into subgroups. The first and second subgroups from the bottom each contain three bands. We have verified that the eigenstates can be nicely attributed to $\psi_{\Pi^*}$ and $\psi_{N}$ as schematically shown in Fig. \ref{fig2}(b). Above the six bands, the $\psi_{\Pi}$ subbands are intertwisted with higher states. As discussed above, without $t_3$ and $t_4$, there does not exist an isolated flat band. Figure \ref{fig3}(c) includes the largest inter-orbital hoppings $t_4=-0.2$ eV, which occurs between the $\beta$ sites [see labels in Fig. \ref{fig2}(a)], and the signature of a flat band as the 7th band begins to show. By further including the largest $t_3$ [Fig. \ref{fig3}(d)], which also occurs between the $\beta$ sites, the flat band is better separated from the other bands above. Finally in Fig. \ref{fig3}(e), when all the smaller (but numerous) inter-orbital hoppings are included, the flat band can be well reproduced. 

As a further comparison, we also list in the first row of Tab. \ref{tab:CDW} the variation range of hopping parameters estimated within the SK two-center approximation as previous adopted~\cite{smith1985band,RossnagelPRB2006}, which scales $dd\sigma$, $dd\pi$ and $dd\delta$ according to the Ta-Ta distance. Since $t_3$ is positive in the $1\times1$ structure, the simple bond length scaling cannot produce a negative value, and $t_4$ is fixed to be zero. Therefore, a large negative $t'$ is absent, which explains why the previous TB calculations failed to reproduce the flat band. The underlying reason is that besides the Ta-Ta distance modulation, CDW also distorts the local TaS$_6$ octahedral symmetry, and in turn modifies the geometry of the WFs (or equivalently, the hybridization between the Ta $d$ and S $p$ orbitals), which is however missing under the Ta-Ta two-center approximation. It is most clear to see from the vanishing $t_4$ in the SK parameterization that the CDW-induced symmetry reduction is not properly taken into account. 

In summary, we present a simple understanding on the band formation physics in 1T-TaS$_2$ by employing a rotated $t_{2g}$ basis. We show that the inter-orbital coupling enhanced by CDW plays an important role in creating an isolated flat band around the Fermi level. The overall framework is also applicable to other structurally similar TMDs with the same CDW pattern, such as TaSe$_2$ \cite{Berkeley} and NbSe$_2$ \cite{NbSe2Qiao}. The quantitative differences of the hopping parameters are expected to result in variations of the electronic structure. By further including inter-layer hoppings, the stacking effects can in principle be included within the rotated basis as well, which presents an interesting problem for future investigation.

This work is supported by the National Natural Science Foundation of China (Grants No. 11774196, 11974197, 51920105002, 12074091), Tsinghua University Initiative Scientific Research Program, and Guangdong Innovative and Entrepreneurial Research Team Program (No. 2017ZT07C341).

\bibliography{ref} 
\end{document}